\documentclass[11pt]{article}
\usepackage[latin1]{inputenc}
\usepackage[english]{babel}
\usepackage[namelimits]{amsmath}
\usepackage{amssymb}
\usepackage{amsmath}
\usepackage{amsthm}

\begin{document}
\title{Statistical ensembles for money and debt}
\author{Stefano Viaggiu\footnote{Corresponding author}
\\
Dipartimento di Matematica, Universit\'a 'Tor Vergata',\\ 
Via della Ricerca Scientifica, 1\\
Rome, Italy 00133,\\
viaggiu@axp.mat.uniroma2.it,\\
Andrea Lionetto\\
Dipartimento di Fisica, Universit\'a 'Tor Vergata',\\
Via della Ricerca Scientifica, 1\\
Rome, Italy 00133,\\
lionetto@roma2.infn.it,\\
Leonardo Bargigli\\
Dipartimento di Scienze Economiche e Sociali,\\ 
Universit\'a Politecnica delle Marche,\\
Piazzale Martelli, 8 \\
Ancona, Italy 60121,  \\
leonardo.bargigli@gmail.com,\\
Michele Longo
\\
Facolt\'a di Economia, Universit\'a Cattolica 'Sacro Cuore'\\
Largo Gemelli, 1 \\
Milano, Italy 20123\\
michele.longo@unicatt.it} 
\date{\today}\maketitle
\begin{abstract}
We build a statistical 
ensemble representation of two economic models describing respectively, in simplified terms, a payment system and a credit market. 
To this purpose we adopt the Boltzmann-Gibbs distribution
where the role of the Hamiltonian is taken by the total money supply (i.e. including money created from debt) 
of a set of interacting economic agents.
As a result, we can read the main thermodynamic quantities in terms of  
monetary ones. In particular, we define for the credit market model a work term which is related to the impact of monetary policy on credit creation. Furthermore, with our formalism we recover 
and extend some results concerning the temperature of an economic system,
previously presented in the literature 
by considering only the monetary base as conserved quantity. Finally, we 
study the statistical ensemble for the Pareto distribution.

\end{abstract}
\section{Introduction}

Historically, the first attempt to establish a link between classical thermodynamics and economics is due to Samuelson \cite{2}, who remained nevertheless quite skeptical about the possibility to introduce concepts borrowed from statistical mechanics, like entropy, into economics\footnote{\textbf{A mapping between classical thermodynamics and economics can be found in \cite{3}, without reference to the underlying statistical structure (see also \cite{3a,3b,3c})}. 
}. This is not surprising, since the neoclassical notion of economic equilibrium, in analogy with classical mechanics, refers to a single point in the phase space, which is the solution of a maximization problem, and not to an equilibrium probability distribution like in the case of statistical mechanics.

When heterodox economists criticize mainstream economic theory by underlining that the economy is not a system in equilibrium, they refer mainly to the first, static, notion of equilibrium. In fact, the general idea of employing statistical equilibrium as a key tool for economics is advocated by some leading heterodox economists (see e. g. \cite{11}), while models borrowed from statistical physics have been successfully applied to microeconomic models where agents are not perfect optimizers and/or not perfectly informed, like for instance in \cite{12} or \cite{13}. On the other hand, unconventional economists are rightly worried about the risk in substitute the unrealistic neoclassical paradigm with an equally unrealistic paradigm derived from statistical physics \cite{14}. In fact, it is quite unreasonable to think of long run economic evolution, which is the outcome of a complex out-of-equilibrium dynamics, as being explained by conservation laws, although we easily admit that such an economic world, described by stable probability distributions, would be indeed one in which our daily life would be much easier.  \\
Some authors \cite{1} argue that if we accept the possibility that 'over a limited period of time, an economic system might behave as though it were quasi equilibrate' in a statistical sense, the knowledge of equilibrium states for economic systems would then explain their possible 'thermodynamical' evolution. Unfortunately, from the perspective of economists this otherwise attractive bargain is based on a claim which is still overreaching, since a national economy is still too complex to behave conservatively even in the very short run. We should be much more modest since, as explained by \cite{14}, the only conservative process in the economic sphere is exchange which, apart from being a short run phenomenon, delivers an inherently partial representation of the economic reality. In particular, equilibrium distributions are conceivable only by separating temporally production and exchange, in such a way that supply is kept fixed over the time span considered in the model. \\
Even taking into account this limitation, we can think of many economic contexts in which the application of the conservation principle could turn out to be useful. In general, conservation may be assumed to hold in all those markets where supply may be taken to change slowly, while exchange takes place at a relatively fast rate, enabling a relaxation of the system towards statistical equilibrium. These markets are quite common in the economic reality. A first simple example is provided by the daily functioning of fish markets, since fishermen cannot expand or shrink their supply of fresh fish once the market is open. A second, perhaps more relevant, example is provided by the real estate market, since the supply of new houses comes necessarily with a significant time lag with respect to positive variations of demand, and a negative adjustment of supply is generally impossible under normal conditions. A third example is provided by financial markets, where the supply of financial assets (stocks, bonds, etc.) can be easily assumed to be fixed on a daily basis. A fourth, less obvious, example is the credit market, since the supply of currency, customary referred to as the monetary base, is controlled by the central bank, and can be assumed to be fixed over a short time span, whereas money supply is determined endogenously by agents interactions through lending and borrowing.\\    
Recently some authors have followed a more ambitious path, by tackling a macroeconomic problem, namely the determination of the most likely data generating process for the empirical distribution of income and wealth at the national level \cite{4,5,6}, with the tools of equilibrium statistical mechanics. In these papers it is argued that the main ingredient
for the derivation of the Boltzmann-Gibbs law is the 
existence of a conserved quantity \cite{7}. 
In this context, the probability distribution 
for an energy value $\epsilon$ is $P(\epsilon)\sim e^{-\epsilon/T}$,
where $T$ is the temperature. The authors of \cite{4,5,6} argue that
money, and not material wealth, could act as a 
conserved quantity in a closed economic system. In fact, material wealth as goods and properties can be consumed 
and destroyed, while an agent (also a firm) cannot print money, but
only exchange it with other agents. 

This assumption has been criticized by 
other authors \cite{15}, who have underlined that money is not a conserved quantity since, as underlined above, the endogenous lending process generally changes the amount of money in the system. From this perspective, the introduction of debt under the form of negative cash \cite{4} clashes with economic realism since cash by definition cannot be negative. In order to be more realistic,  we need our definition of money to be consistent with the well established standard notions of monetary base and money supply. In order to keep terminology as simple as possible, in this paper we choose to define money (cash), unless otherwise specified, as synonym of the monetary base, while the sum of currency and of highly liquid credit assets, like demand deposits (which we will indicate simply as 'credit'), will be labeled as total money supply.   

Furthermore, in \cite{4} a Boltzmann-Gibbs equilibrium distribution for money is obtained as a result of numerical agent-based simulations. 
A similar study, with an attempt to introduce 
an ensemble for the conservation of money, is given in
\cite{10}. In fact, the introduction of a Boltzmann-Gibbs distribution,
in order to build a statistical mechanics for money, should also
be justified by introducing an ensemble starting from the microscopic 
economic variables
from which total money supply is composed, similarly to the introduction 
of the coordinates for a given ensemble.
In this way, the introduction of aggregate economic variables 
is obtained after an integration over the microscopic variables 
which describe the total money supply of a given economic system. 

This approach is the starting point of our paper. We will see that this representation adapts naturally to economic interpretation. In particular, with this formalism it is possible to introduce credit as a further variable 
with respect to the conserved monetary base, providing results which are consistent with the endogeneity of total money supply required for a sound economic interpretation.

The plan of the paper is the following:
in the next section we introduce our formalism together 
with its economic interpretation, 
in section 3,4,5 we introduce the statistical ensembles for money,
in section 6 we analyze some macroeconomic relations suggested by 
the thermodynamics analogy while in section 7 we track some conclusions 
and final remarks.
The appendix is devoted to the application of our method to the Pareto law.

\section{Preliminaries}
In this paper we want to introduce sound statistical ensembles for economical systems 
with money (cash) and credit-debt. 
First of all, in our model the $N$ particles of the standard physical
approach are substituted by $N$ 
interacting economical agents. In this context,
we use the coordinate $x_i$ to label the cash (money) possessed by the
$i_{th}$ agent, while $y_i$ represents its credit-debt variable.  
In our context, we have no need to employ the complete Hamiltonian formalism of the ordinary
statistical mechanics. In particular, the variables $\{x_i,y_i\}$ we use are not conjugated in
the sense of the Hamiltonian mechanics. 
To build an ensemble we have only need of a conservation law. We take as a conserved quantity the total money supply function $M$,
which can be considered constant over a suitable time interval.
Moreover, the function $M$ will be a function
of the chosen coordinates, i.e. $M=M(x_i,y_i)$.\\
At this point the ergodic hypothesis comes into action which permits us,
given a function $f(x_i,y_i)$, 
to express its average with respect to the time in terms of an average over the ensemble
at fixed $M$: 
\begin{equation}
\overline{f}=\int_{M=const}f(x,y)\rho(x,y)dx\;dy,
\label{r1}
\end{equation}
where $\rho (x,y)$ denotes the probability distribution of the ensemble. 
Taking into account these considerations, our fundamental aim in this paper is 
to show that ensembles can be introduced in a consistent way for an economic system. 
To this purpose, we study simple money functions reproducing known results present in the literature, such as the following:
\begin{equation}
M=\displaystyle\sum\limits_{i=1}^N {(x_i+y_i)}
\label{master}
\end{equation}
In expression (\ref{master}) both $x_i$ and $y_i$ can be expressed in monetary units.
We stress that the expression (\ref{master}) is the simplest choice we can consider. 
More sophisticated models can be introduced by means of an interacting term $I(x,y)$ in
the money supply function $M$. Once the ensembles have been defined, we can introduce in a consistent way all the thermodynamic functions in terms of $N$ and of economic variables, and finally we can study economic (thermodynamical) transformations. 
\section{Microcanonical ensemble}
In order to introduce the microcanonical ensemble we consider an isolated economic system with $N$
agents, where money $M$ is fixed. As in the usual microcanonical ensemble in statistical mechanics,
if we integrate over all the available volume of the configuration space spanned by the variables
$\{x,y\}$ with 
$\overline{M}=m=constant$
(the overline denotes the average over the whole configuration space), 
then we have $\Omega=\int_{\overline{M}=m}d^N x\;d^N y=0$ (see \cite{huang}
and references therein).
As usual in statistical mechanics, we introduce a thick shell $\Delta$ where $\Delta <<m$ and define
\begin{equation}
\Gamma(m)=\int_{m<M<m+\Delta} \frac{d^N x\;d^N y}{k^{2N}},
\label{4}
\end{equation}
where $k$ is a normalization factor such that $\Gamma$ is dimensionless.
In the following we set $k=1$.
The average of a given quantity $\overline{f}$ is given by
\begin{equation}
\overline{f}=\frac{\int fd\Gamma}{\Gamma}.
\label{5}
\end{equation}
To calculate $\Gamma(m)$ it is convenient to introduce the following integral:
\begin{equation}
\Sigma(m)=\int_{M\leq m} d^N x\;\;d^N y,
\label{6}
\end{equation}
where 
\begin{equation}
\Gamma(m)=\Sigma(m+\Delta)-\Sigma(m)\simeq\frac{\partial\Sigma(m)}{\partial m}\Delta.
\label{7}
\end{equation}
The functional $\Gamma(m)$ measures the number of microscopic realizations of a given
economic system. The entropy $S$ can be defined in the usual way:
\begin{equation}
S=\ln\Gamma(m)=\ln\Sigma(m),
\label{8}
\end{equation}
where the last equality in (\ref{8}) follows for $N>>1$ (see \cite{huang}).
In our description of an economic system, entropy keeps its ordinary statistical meaning, i.e. it is proportional to the number of micro configurations of the system, which is consistent with a given distribution of the $N$ agents with respect to $x$ and $y$. Then, by analogy with ordinary statistical systems, the
equilibrium configuration is the one that maximizes the entropy. This condition,
together with the conservation of money (energy) leads to the ensemble distribution.\\ 
After writing $S=S(m,V)$, where $V_y=\int dy$, we can define the analogues in our context
of temperature $T$ and pressure $P$:
\begin{eqnarray}
& &dS=\frac{\partial S}{\partial m}dm+\frac{\partial S}{\partial V}dV,\label{9}\\
& &{\left(\frac{\partial S}{\partial V}\right)}_{m,N}=\frac{P}{T},\;\;
{\left(\frac{\partial S}{\partial m}\right)}_{V,N}=\frac{1}{T}. \label{10}
\end{eqnarray}
Furthermore, we define the analogue of the free Helmholtz energy $F$:
\begin{eqnarray}
& &F=m-TS,\label{11}\\
& &dF=dm-TdS-SdT.\label{12}
\end{eqnarray}
From (\ref{9}) and (\ref{10}) we get:
\begin{equation}
TdS=dm+PdV. \label{13}
\end{equation}
Equation (\ref{13}) is the analogue of the first thermodynamic principle.
From (\ref{12}) and (\ref{13}) we have:
\begin{eqnarray}
& &dF=-PdV-SdT,\label{14}\\
& &{\left(\frac{\partial F}{\partial V}\right)}_{T,N}=-P,\;\;
{\left(\frac{\partial F}{\partial T}\right)}_{V,N}=-S.\label{15}
\end{eqnarray}
In the following section we give an economic interpretation of the quantities $P$ and $T$, while $V_{y}$ is naturally interpreted as the maximal amount of credit/debt per agent allowed in the system.\\ 
As a first warm-up example, we want to reproduce known result in the literature \cite{4},
\begin{equation}
M=\sum\limits_{i=1}^N x_i
\label{linM}
\end{equation}
In the expression (\ref{linM}) we have supposed one component for any pair of coordinates
$(x,y)$, with $x\in[0,\infty)$. 
This is the simplest assumption we can make. Integration of (\ref{6}) is easy and 
it gives:
\begin{equation}
\Sigma(m)=\frac{m^N}{N!}V_y^N,
\label{ii}
\end{equation}
By using (\ref{8}) and (\ref{10}) we have:
\begin{equation}
S = N (\ln m + \ln V_{y}) - \ln N!,\;T=\frac{m}{N},\;
P=\frac{m}{V_{y}}=\frac{NT}{V_{y}}.
\label{iii}
\end{equation}
The expression for $T$ of (\ref{iii}) is the same as that of the model of Dragulescu and 
Yakovenko \cite{4}.
In the form 
(\ref{iii}), $T$ measures the mean money per agent.
In the following section we obtain different models in terms of the more usual
canonical ensemble.

\section{Canonical ensemble}

As the usual canonical ensemble, we have an economic system with $N_1$ agents
with money $m_1$ interacting with another system with $N_2$ agents (with $N_2>>N_1$) 
with money
$m_2$. 
Under the assumptions that total money $m=m_1+m_2$ is conserved and that:
\begin{equation}
\Gamma(m)\simeq {\Gamma}_1(m_1){\Gamma}_2(m_2=m-m_1),
\label{20}
\end{equation}
we can  calculate the probability distribution at equilibrium of the subsystem
$1$ independently of the behavior of reservoir $2$.
As a consequence, for an economic system with $N$ agents, the money function $M$
and economic temperature $T$ are given by
\begin{eqnarray}
& &dP=\frac{d^N x\;d^N y\, e^{-\frac{M}{T}}}{Z},\label{21}\\
& &Z=\int d^N x\;d^N y\, e^{-\frac{M}{T}},\label{22}
\end{eqnarray}
Moreover, by a simple algebra we obtain
\begin{equation}
Z=e^{-\frac{F}{T}},
\label{23}
\end{equation}
with $F$ given by (\ref{11}).
It is natural to introduce the analogue of the Maxwell-Boltzmann
distribution for a single agent.
Actually the infinitesimal probability 
$dP_{i}$
that the $i_{th}$ agent is in a point of its configurational space
$\{x,y\}$ independently of the behavior of the other agents
can be written as
\begin{equation}
dP_{i}(x_i,y_i)=ce^{-\frac{m_i}{T}}dx_i dy_i, 
\label{24}
\end{equation}
where $c$ is a normalization constant.
Note that also in \cite{4,5,6} a distribution
similar to (\ref{24}) is considered, but integrated in a different space. In fact,
in \cite{4,5,6} the exponential in (\ref{24}) is integrated 
with respect to $dm_i$, while in (\ref{24})
according to the definition of the Boltzmann-Maxwell distribution for a classical ideal gas, 
it is integrated with respect to
$\{x_i,y_i\}$. 

\subsection{Model of a payment system with only cash}
We first derive the model (\ref{linM}) in terms of the usual Lagrange multipliers (LM) by a 
discretization of the system and after we translate the model in terms of ensembles.
The model describes the following economic situation. The generic agent $j$
with initial cash $x_j^I$ pays a certain amount of money $\Delta x$ to another agent 
$k$ with initial cash $x_k^I$. In the final configuration we have
the agent $j$ with cash $x_j^{II}=x_j^{I}-\Delta x$ and the agent
$k$ with final cash $x_k^{II}=x_k^{I}+\Delta x$ with the sum conserved. This situation, as underlined by \cite{b4}, describes a payment system, which is related to a unobservable flow of goods and services across agents.\\
The number of configurations $\Gamma$ of $n_k$ agents with money $m_k$ is
\begin{equation}
\Gamma(\{n_k\})=\left[\frac{N!}{n_1!n_2!\cdots n_k!}\right].
\label{r2}
\end{equation}
The entropy $S=\ln\Gamma(\{n_k\})$ must be maximized with the LM
$\alpha$ and $\beta$ and with the constraints
\begin{equation}
\sum_p n(p)=N,\;\;\;\sum_x n(x) x = m = \left<M\right>,
\label{r3}
\end{equation}
by taking 
\begin{equation}
\overline{S}=S-\alpha\sum_x n(x)-\beta\sum_x n(x) x.
\label{r4}
\end{equation}
The maximization of (\ref{r4}) gives $n(x)=e^{-\alpha-\beta x}$.
By introducing this expression in (\ref{r3}) and changing summation with integration,
i.e. $\sum n(x)\rightarrow\int n(x)dx\;dy$, we obtain
$1/\beta=T=m/N$.\\
Within the canonical ensemble we have
\begin{equation}
Z=\int d^Nx\;d^Ny e^{-\frac{\sum\limits_{i=1}^N x_i}{T}}.
\label{r5}
\end{equation}
We easily obtain, by a factorization of (\ref{r5}), $Z=V_y^N T^N$. By means of
(\ref{23}) we have $F=-T\ln Z$. By means of (\ref{15}) and (\ref{11}) we obtain 
for $T$ the expression $T=m/N$. This case is similar to the perfect gas one of the usual 
statistical mechanics. The gas is in a box of volume $V_{y}$. As stated above, 
in the economic system, 
$V_y$ can be interpreted as the maximal credit level allowed in the system, although in this case  
the function
$M$ does not contain explicitly credit-debt terms.
Any 
thermodynamical economical transformation which increases $V_y$ simply increases the 
allowed credit level.

\subsection{Models of a payment system with debt}
In the model \cite{4} debt is introduced by allowing for negative money.
In our approach money is always positive, and as a consequence our analogue
of \cite{4} is given by $M=\sum\limits_{i=1}^N y_i$, 
where $y_i\in[-d,\infty), d\geq0$.
The economic dynamics of this model is the same as that of the previous subsection above, 
but $y$ here represents a current account balance with maximal overdraft $-d$. Then, if $y_{i} > 0$ agent $i$ has a credit claim towards a bank, supposed to be external to the system, which he can use to pay for goods and services, while if $y_{i} < 0$ the same agent is borrowing from the same bank in order to fulfill the same payments.
In terms of the LM technique we have 
$\sum n(y)=N$ and $\sum n(y) y = Q_0=\left<M\right>$.
By the same technique as that of the previous section , with 
$\sum n(y)\rightarrow\int n(y)dy\;dx$ and  $y\in[-d,\infty)$ we obtain  
\begin{equation}
T=\frac{Q_0}{N}+d. 
\label{r6}
\end{equation}
In terms of our formalism we have 
\begin{equation}
Z=\int d^Nx\;d^Ny\;e^{-\frac{\sum\limits_{i=1}^N y_i}{T}}.
\label{r7}
\end{equation}
By performing the same calculations as that of the previous subsection  we get
$Z=V_x^N T^N e^{\frac{Nd}{T}}$. As a result we obtain for $T$ the expression (\ref{r6}).
In this case $V_x$ represents the maximal amount of cash allowed in the system.
{Note that we can also have $Q_0<0$, with the obvious condition
$T\geq 0$, i.e. $d\geq -Q_0/N$. Moreover, we can build a model
with $d\leq0$ and the formula (\ref{r2}) is again valid.
The model (\ref{r7}) can be generalized further. In fact, suppose that each agent $i$ 
has $r_{i}$ different current accounts $\left\lbrace y_{i1},y_{i2} \dots y_{ir_{i}} \right\rbrace $ with maximal overdrafts $\left\lbrace d_{i1},d_{i2} \dots d_{ir_{i}} \right\rbrace $. Then we have
\begin{equation}
T=\frac{Q_{0}}{R} + \frac{1}{R}
\sum\limits_{i,j=1}^{N,r_{i}}d_{ij}
\label{r8} = \left \langle y \right \rangle + \left \langle d \right \rangle
\end{equation}
where $R =  \sum_{i = 1}^{N} r_{i}$.
We now consider the more general money function given by (\ref{master}), which represents a system where agents make payments either by cash or through a single current account (e.g. by using bank checks or credit cards). It should be noticed that in this case the constraint is on $m_{i} = x_i+y_i$, since we suppose that agents can freely convert $x_i$ into $y_i$ and \textit{vice versa}. As a result,
no constraints are imposed on $\sum y_i$ or $\sum x_i$. \\
In the first place, we can build a model where both credit and debt are allowed.
In terms of LM we have 
$\sum n(m)=N,\;\sum n(m) m = \left<M\right>$ where 
$y_i\in[-d,\infty), d\geq0$. By means of the same technique we previously used
(remember that in this case $\sum n(m)\rightarrow\int n(m)dx\;dy$
with (\ref{r2}) again valid) we obtain
\begin{equation}
T=\frac{1}{2}\left(\frac{m}{N}+d\right).
\label{r9}
\end{equation}
In order to translate the model in terms of the ensemble formalism,
we proceed exactly as in subsections 4.1 and 4.2.
As a result of the integration of (\ref{22}) and thanks to (\ref{23}), (\ref{11}) and (\ref{15}) we obtain
\begin{eqnarray}
F&=&-2NT\ln(T)-N d\label{i1}\\
S&=&2N\ln(T)+2N\label{i1bis}\\
T&=&\frac{1}{2}\left(\frac{m}{N}+d\right).\label{i2}
\end{eqnarray}
We remark that the temperature we get is one half the one obtained
in \cite{4},
This factor is related to the two alternative payment means for each agent we have in our model. In fact, it's easy to see that (\ref{i2}) is a special case of (\ref{r8}) with   $y_{i1} = x_{i}$, $y_{i2} = y_{i}$, $r_{i} = \text{const} = 2$, $d_{x} = 0$ and $d_{y} = \text{const} = d$. 
Thus, the introduction of alternative payments means for a fixed $m$ makes the tail of the Boltzmann distribution for money thinner, by decreasing the value of $T$.

We now analyze a model, defined by the same money function (\ref{master}), where credit is not allowed, i.e. $y_i\in[-d,0]$ . This model is consistent, although not necessarily realistic, if we allow total cash to be increased by debt creation.
To this purpose, consider an initial 
condition in which a generic agent '$1$' has $x_1^0=x_1,\;y_1^0=0$ and another 
labeled '$2$' has $x_2^0=x_2,\;y_2^0=0$. In a first phase 'I' the agent '$2$' borrows money
from '$1$'. We have $x_1^I=y_1-\Delta x,\;y_1^I=\Delta x=\Delta y$ and
$x_2^I=x_2+\Delta x,\;y_2^I=-\Delta y$. Moreover, the agent '$1$' can, by means of a central bank,
change credit with respect to the agent '$2$' with money cash. Hence, in the final 
state 'II' we have $x_1^{II}=x_1-\Delta x+\Delta x=x_1,\;y_1^{II}=0$ and 
$x_2^{II}=x_2^I,\;y_2^{II}=y_2^I$. In this way, the total amount of cash is increased, since $x_1^{II}+x_2^{II}=x_1+x_2+\Delta x$, with a net creation of debt $-\Delta x=\Delta y$. 
This result can be interpreted as a debt of the agent with a central bank.
This model can be solved by integrating with $y_i\in[-d,0]$.\\
By integrating equation (\ref{r5}) for $Z$ and with the help of (\ref{23}), (\ref{11}) and (\ref{15}) we get:
\begin{eqnarray}
& &F=-2NT\ln(T)-NT\ln\left(e^{\frac{d}{T}}-1\right),\label{i3}\\
& &S=2N\ln(T)+2N+N\ln\left(e^{\frac{d}{T}}-1\right)-
\frac{Nd\;e^{\frac{d}{T}}}{T\left(e^{\frac{d}{T}}-1\right)},\label{i4}\\
& &m=2NT-\frac{Nd\;e^{\frac{d}{T}}}{\left(e^{\frac{d}{T}}-1\right)}.\label{i5}
\end{eqnarray}
Note that the expression $m$ in (\ref{i5}) is much more involved than the one of \cite{4}.
However, for $T>>d$ expression (\ref{i5}) for the temperature $T$ becomes
\begin{equation}
T=\frac{m}{N}+\frac{d}{2}+o(d),\label{i6}
\end{equation}
which is a result which differs, at the given approximation order, from that in \cite{4} only in the coefficient $1/2$ (instead of $1$) of the linear term $d$. 

\subsection{Model of a credit market}

A credit market differs from a payment system in that it is constrained to respect the basic accounting identity over interactions between pairs of agents. If the accounting of a credit relationship is correctly reproduced, when agent 1 lends to agent 2 we have
\begin{eqnarray}
& &M_{1} = (x_{1} - \Delta m) + (d_{1} + \Delta m) = \text{const}_{1}\\
& &M_{2} = (x_{2} + \Delta m) + (d_{2} - \Delta m) = \text{const}_{2}\label{ecce}
\end{eqnarray}
In this case, $d_{i}$ stands for net debt, i.e. $d_{i} = \text{assets}_{i} -  \text{liabilities}_{i}$. In our case, we have a single class of assets (credit) and of liabilities (debt) respectively. Furthermore, the following additional constraints must hold:
\begin{eqnarray}
 &\sum  x_i  = M_{0} \\
 &\sum  \text{assets}_{i}  - \sum \text{liabilities}_{i}  = Q_{0}
\end{eqnarray}
The first constraint means that the monetary base $M_{0}$ is fixed, i.e. that agents, by exchanging credit, can increase or decrease only the overall money supply (i. e. the sum of money (cash) and credit), while $M_{0}$ remains fixed. The second constraint means that the system is closed, i.e. any credit claim of some agent $i$ in the system represents a debt for some other agent $j$ in the system and vice versa. In the following, we suppose without loss of generality that $Q_{0} = 0$, so that credit and debt distributions are identical.
The money function for this model is given by $\sum y_i = m$, where $y_{i}$ stands for credit, i.e. $y_{i} \in [0,+\infty]$. In fact, since $M_{0}$ is fixed, the only source of fluctuations for $M$ in the canonical ensemble is given by credit/debt $\sum_{i}^{N} y_{i}$. This fact allows us to obtain a well defined temperature for the system.
In terms of the ensemble formalism we have the following:
\begin{equation}
Z = \int d^Nx\;d^Ny\;e^{-\frac{M}{T}} = \int d^Nx\;d^Ny\;e^{-\frac{\sum\limits_{i=1}^N y_{i}}{T}} =  M_{0}^{N} T^{N},
\label{4.3.1}
\end{equation}
where in (\ref{4.3.1}) we have posed $M_0=V_x$. This assumption can be understood in the following way. Since $M_0$ is fixed, we have that $\sum x_i=M_0$, and changing summation with integration we have ${\int}_{\sum x_i\leq M_0}dx=V_x=M_0$. 
Furthermore, it is easy to see that the expression (\ref{iii}) for $P$ still holds, with $V = V_x = M_0$.
By following the same steps as the previous section we obtain again
\begin{equation}
T=\frac{m}{N},
\label{ad1}
\end{equation}
Of course, we could obtain the same expression by integrating only over the $y_{i} \in [0,+\infty]$. Nevertheless, we maintain the usual representation in terms of $x_{i}$ and $y_{i}$ in order to reproduce correctly the main thermodynamical equations. In fact, as we will see, the variations of $\sum x_{i} = M_{0}$ as determined by monetary policy are needed in order to introduce a ``work'' term in the variation of credit. In this context, the variations of $m$ reflect instead a change of attitude on the part of the agents, which may become more or less oriented towards credit creation. \\ 
Finally, we introduce a generalization of the money function (\ref{master}) of the following kind:
\begin{equation}
M=\sum\limits_{i=1}^N \sum\limits_{j=1}^I y_{ij}
\label{rr1}  
\end{equation}
where the index $i$ runs over the agents while $j$ runs over the different asset classes allowed in the model. This generalization turns out to be useful to describe a full fledged financial market with different kinds of assets and liabilities. In this case we obtain the following
\begin{eqnarray}
F&=&-TNI \ln T\\
S&=&NI \ln T + NI \\
T&=&\frac{m}{NI}
\end{eqnarray}
Thus, if one allows more kinds of assets/liabilities into the system, the likelihood of observing large values of $m_{i}$, as well as large deviations from $m$, will be lower, since $T$ is lower for the same $m$ with respect to the one-dimensional case with temperature (\ref{ad1}). In other words, diversification makes the system more stable, a result which is consistent with standard economic theory. 
For a further discussion on debt
the reader can refer also to \cite{b4,b5}. In particular, in the model present in \cite{b5} a constraint on debt is introduced by adopting the fractional reserve system of banking. Even this model could be reproduced with our formalism in terms of a statistical ensemble.

\section{Grand Canonical Ensemble}

This kind of ensemble is aimed at describing the probability distribution 
of a system in which the agent number is not conserved.
The probability distribution with $N$ agents is given by
\begin{equation}
dP_{rN}=e^{\left(-\frac{PV}{T}+\frac{\mu N}{T}-\frac{M}{T}\right)}
d^N x\;d^N y,
\label{25}
\end{equation}
where $\mu$ is the analogue of the chemical potential and:
\begin{eqnarray}
& &dS=\frac{dm}{T}+\frac{P}{T}dV-\frac{\mu}{T}dN,\label{26}\\
& &dF(T,V,N)=-PdV-SdT+\mu dN,\label{27}\\
& &{\left(\frac{\partial F}{\partial N}\right)}_{T,V}=\mu=
{\left(\frac{\partial m}{\partial N}\right)}_{S,V}. \label{28}
\end{eqnarray}
The last equality in (\ref{28}) follows from the Maxwell relations \cite{7}.
From (\ref{26}) we have:
\begin{equation}
TdS=dm+PdV-\mu dN.
\label{29}
\end{equation}
The analogue of the Gibbs-Duhem relation gives
\begin{equation}
SdT=VdP-Nd\mu.\label{30}
\end{equation}
In this context, the 'potential' term $\mu$ takes into account the contribution to
$m$ caused by a variation of the
number of the agents. For an interpretation of $\mu$ in terms of immigration see
\cite{b2}.

\section{Thermodynamics of Money: Some Relations}
The starting point for a thermodynamics of money
is provided by equation (\ref{29}). In the following, we will refer to the credit market model as a benchmark to provide a sound economic interpretation of thermodynamical relationships. We should remember that in this model $V  = V_{x}$, i.e. $V$ is related to the cash variable $x$ and thus to the monetary base. First of all,
thanks to (\ref{28}) and by taking $(S,V)=constant$ we get
\begin{equation}
dm={\left(\frac{\partial m}{\partial N}\right)}_{S,V}dN.
\label{31}
\end{equation}
Hence, the analogue of the thermodynamic chemical 
potential in our context has to be identified with the variation of credit induced by a variation of the agent number.
In complete analogy with thermodynamics we set $TdS=dC$ in (\ref{29}), 
where $dC$ represents the increase (or decrease) of credit, which is caused respectively by a change in the attitude of agents ($dm$), by a change of the monetary base induced by the central bank($PdV$), or by a change in the number of agents ($\mu dN$). In fact,
further pursuing the analogy with thermodynamics 
we can compute the quantity
\begin{equation}
L_{i\rightarrow f}=\int_{i\rightarrow f} PdV, \label{32}
\end{equation}
where $i,f$ refer to an initial and a final economic state
characterized by $(T,P,V,N)$. The integral (\ref{32}) depends in general on the path
from $i$ to $f$. 
Moreover, as underlined, $V=\int dx$ is a measure of the monetary base present in the system.
Then a change in $V$ represents a change of monetary policy by the central bank, and it can be naturally interpreted as the work done on the system by the central bank itself.
Summarizing, for quasi-static economic transformations, we can write
\begin{equation}
dC=dm+dL-\mu dN,
\label{33}
\end{equation}
In this way, equation (\ref{31}) can be rewritten as:
\begin{equation}
dm={\left(\frac{\partial m}{\partial N}\right)}_{L,C}dN.
\label{31bi}
\end{equation}
Furthermore, if the economic transformations 
can be considered as quasi-static, then we can insert in (\ref{32}) the expression 
(\ref{iii}) for $P$  which is valid for a 'free' agent. In our case, $P = \frac{m}{V_{x}}$, i.e. $P$ in our context captures the relationship between credit and money, by describing the magnitude of credit for a given value of the monetary base. We remark that, if we followed the fractional reserve system model of \cite{b5}, we should write $m = \left(\dfrac{1}{r}-1 \right) V$, where $r$ is again controlled by the central bank. In this case, credit would be completely determined by the central bank itself.\\
If we consider more general transformations than quasi-static ones, equation
(\ref{33}) becomes:
\begin{equation}
\delta C=dm+\delta L-\mu dN,
\label{34}
\end{equation}
where the non exact differential $\delta$ takes the place of the exact one $d$.
In this way we can consider economic transformation that are not
'quasi-static'. As an example we can study the analogue of an irreversible
transformation in the usual thermodynamics by considering an isolated 
economic system evolving 'spontaneously'. In this case, as the system is isolated,
$(C,L,N)$ are left unchanged. Hence also $dm=0$, and the economic 
temperature remains unchanged (as it happens for a free expansion of an isolated gas),
provided that the money function $M$ contains only 'free' terms $y_i$. 
We can also calculate the entropy $S$ as:
\begin{equation}
S_f-S_i=\int_{i\rightarrow f}\frac{\delta C}{T},
\label{35}
\end{equation}
where the integral (\ref{35}) must be calculated over any
quasi-static transformation connecting the initial and the final economic states.
Note that the formula (\ref{34}), for a change with fixed $C$ and agent number
$N$, yields:
\begin{equation}
dm+\delta L=0,
\label{33b}
\end{equation}
Adiabatic transformations are then naturally interpreted as those where the central bank counteracts the endogenous increase (decrease) of credit with an opposite variation of the monetary base in such a way that the overall credit remains fixed. \\
We have seen that in the credit market model it is possible to define a 'work term', contrary to what happens in the payment system models, as underlined in \cite{b2}. In order to show that the introduction of this term is consistent in this model, we define the monetary analogue of a Carnot heat engine, and show that the derived cycle has a consistent interpretation in terms of monetary policy.\\
The monetary Carnot engine is ideal since isothermal transformations require that the central bank is able to expand the monetary base without minimally affecting agents attitude regarding credit creation. For this reason, the monetary Carnot engine, as in thermodynamics, provides an upper bound for the effectiveness of monetary policy.
This argument becomes apparent in the fractional reserve model where, taking into account (\ref{ad1}), for isothermal transformations we have:
\begin{eqnarray}
T =  \left ( \frac{1}{r} - 1 \right ) \dfrac{V}{N} &=& \left ( \frac{1}{r'} - 1 \right ) \dfrac{V'}{N} \\
V' - V &=&  \dfrac{V'}{r'} - \dfrac{V}{r}
\end{eqnarray}
The r.h.s. of the last equation represents the variation of money supply, which is set to be equal to the variation of the monetary base. 
\\
As underlined above, in adiabatic expansions an increase (decrease) of credit is associated with a decrease (increase) of the monetary base. In the case of real world monetary policy, this is a reasonable outcome. In fact, monetary easing is usually associated to a contraction of credit and vice versa, since the former is a reaction to the latter. \\
In the monetary Carnot engine we obtain for $L$, with 
temperatures $T_c,T_h,\;T_h>T_c$, the following expression for $L$
\begin{equation}
L = (T_h - T_c) (S' - S). 
\end{equation}
In thermodynamics the performance parameter 
$\eta$ of an engine working between two sources with 
temperatures $T_c,T_h,\;T_h>T_c$
is defined as 
\begin{equation}
\eta=\frac{L}{|Q_h|}, 
\end{equation} 
where $L$ is the work
made by the engine and $|Q_h|$ is the heat absorbed from the source at temperature 
$T_h$. For a Carnot engine we have
\begin{equation}
1-\frac{|Q_c|}{|Q_h|}\leq 1-\frac{T_c}{T_h}=\eta,
\label{36}
\end{equation}
where the equality holds for quasi-statical transformations. The 
Carnot theorem states that the performance $\eta$ of a given heat engine working between two temperatures $T_h,T_c$
can never exceed the one given by a Carnot engine working with only
reversible transformations. In our credit market model, $T_h, T_c$ define two levels of average credit, with the policy performance factor $\eta$ given by
\begin{equation}
\eta=\frac{L}{|C_h|},
\label{37}
\end{equation}
where $L$ is the work done by the central bank through monetary policy 
and $|C_h|$ is the overall credit variation. The policy performance factor can be interpreted as as measure of the effectiveness of monetary policy. By maximizing $\eta$, in other words, we maximize the impact of monetary policy over credit creation. As explained above, this impact is maximal if the central bank is able to follow the monetary Carnot cycle.
Hence, instead of (\ref{36}) we have:
\begin{equation}
1-\frac{|C_c|}{|C_h|}\leq 1-\frac{T_c}{T_h}\;
\rightarrow\;\frac{|C_c|}{|C_h|}\geq\frac{T_c}{T_h}.
\label{38}
\end{equation}
For a constant number of agents we get:
\begin{equation}
\frac{|C_c|}{|C_h|}\geq\frac{m_c}{m_h}
\label{388}
\end{equation}
Equation (\ref{388}) is an inequality relating measurable economic quantities. In particular, it states that the evolution of market attitudes, as given by the r.h.s. ratio, provides an upper bound for the effectiveness of monetary policy.

\section{Results and final remarks}

In this paper we built ab initio the statistical
ensemble representation of two simple economic models. The starting point was the
assumption of the conservation of money, defined as equivalent to the monetary base,
for a system composed by a given number of agents \cite{4,5,6}.
By using standard statistical techniques,
we wrote the economic analogue of all the main formulas
of statistical mechanics where the role of the Hamiltonian 
was played by the money supply function $M$. In our context, the complete Hamiltonian formalism of the 
ordinary statistical mechanics is not necessary. We only need the fact that
there exists a function (the money supply function) which can be considered constant over a
suitable time interval. In fact, 
the coordinates we use, i.e $(x,y)$, are not conjugate.
The analogy between the thermodynamics of a physical system
and that of a credit market is 
based on a sound economic principle, i.e. the agents cannot 'print'
but only exchange money. Only a central bank can print money.
Once the identification between 
thermodynamical quantities and economic ones is performed, we can take into account 
'economic' transformations in which money can be exchanged between
different economic systems or work can be performed under the form of a credit expansion induced by monetary policy.\\  
Although real economic systems are typically not at equilibrium, 
this approach can be used in principle whenever we can introduce a given economic conserved quantity.
In particular, the study of equilibrium configurations can be useful to investigate 
the spontaneous evolution of those economic subsystems in which the conservation principle may be reasonably assumed as valid. These are particular examples of markets in which, with some approximation, supply can be taken as fixed. As we have argued in the introduction, the credit market is an instance of this category inasmuch as the monetary base is determined exogenously by the central bank.\\    
In the context of our formalism we recovered the result of Dragulescu et al. \cite{4,5,6} 
for a system described by the money function (\ref{linM}). In this case we obtain the same result
for the economic temperature present in \cite{4},
with and without the presence of debt (see formula (\ref{r6}).
We stress that this result is obtained as a computation performed by employing the
well-known mechanical statistic approach for the microcanonical ensemble, while the result of \cite{4} is obtained by fitting with an exponential Boltzmann distribution the result of a numerical simulation. 
Moreover we believe it is worth noticing the linear dependence on the variables $x_i$ of the money function. This is the most striking difference from the thermodynamics of a physical system which involves a kinetic term with depends quadratically on $x_i$.     
We extended our analysis by considering generalization of the money function given by (\ref{master})
by introducing the variables $y_i$ whose aim is to describe agents with different classes of assets and liabilities. 
We considered different models, without constraint on $\sum x_i$ or $\sum y_i$,
 whose results are conveniently summarized in table (\ref{Msystems}) where we report $M$ and $T$, which have different interpretations in the different models presented, and the presence of debt and/or credit. 
\begin{table}[t]
\begin{center}
\begin{tabular}{|c|c|c|c|}
\hline
$M$ & $T$ & debt & credit \\
\hline
$\sum\limits_{i=1}^N x_i$ & $\frac{m}{N}$ & $0$ & $0$ \\
$\sum\limits_{i=1}^N y_i$ & $\frac{Q_0}{N}+d$ & $d$ & $\infty$ \\
$\sum\limits_{i=1}^N (x_i+y_i)$ & $\frac{1}{2}\left(\frac{m}{N}+d\right)$ & $d$ & $\infty$ \\
$\sum\limits_{i=1}^N (x_i+y_i)$ & $\frac{m}{2N}$ & $0$ & $\infty$ \\
$\sum\limits_{i=1}^N (x_i+y_i)$ & $\frac{m}{N}+\frac{d}{2}+o(d)$ & $d$ & $0$ \\
\hline
\end{tabular} 
\caption{Temperature $T$ of the economic systems described by the money function $M$ and by the allowed credit and debt.}
\label{Msystems} 
\end{center}
\end{table} 
 As a final application of our formalism, we explicitly show in the appendix which money function $M$ is needed in order to recover a power law Pareto distribution. For a discussion on this issue see also
\cite{b3}.
We leave for future work the application of our approach to other models from which different probability distributions are derived (see \cite{16} for a review).
 
\section*{Acknowledgements}
A L acknowledges V. M. Yakovenko for having clarified some details of \cite{4}.
L B acknowledges the financial support from the European Community
Seventh Framework Programme (FP7/2007-2013) under Socio-economic
Sciences and Humanities, grant agreement no. 255987 (FOC-II).

\section*{Appendix}
In \cite{1} it has been shown how if the following law holds
\begin{equation}
\frac{dI_i}{dt}={\alpha}_i I_i,
\label{39}
\end{equation}
for an agent $i$ with income or wealth $I_i$,
under the hypothesis that ${\alpha}_i$ is randomly distributed around zero,
then there exists a conserved quantity
$Y=\displaystyle\sum\limits_{i=1}^N\ln(I_i)$.
As a consequence, the Pareto power law \cite{102} arises.
Thus we can use the formalism developed in this paper to study the
statistical mechanics and thermodynamics which reproduces the Pareto law for the income. 
To this aim we used the canonical ensemble to describe the microscopic states of the system. 
The calculations are similar to the ones of the previous sections but
now the conserved quantity is $Y$. In this case instead of (\ref{22}) we get
\begin{equation}
Z=\int\frac{d^N x\;d^N y}{k^{2N} N!}{\prod}_i{\left(\frac{T_{max}}{I_i}\right)}^{\frac{T_{max}}{T}},
\label{40}
\end{equation}
where $T$ is the economic temperature relative to the conserved quantity $Y$
and $T_{max}$ is interpreted, as we see below, as the maximum for the temperature allowed 
by the system. 
The integral (\ref{40}) is adimensional.  
As a result, we have a relation involving $T,N$ and
${\overline{Y}}$, i.e. the mean value of $Y$. As an example, we consider again
a 'free' income agent function where $I_i=x_i$.\\
Similarly to (\ref{23}), we have 
$Z=e^{-\frac{F}{T}},\;F=\overline{Y}-TS$ ($S$ is the entropy function)
and ${\left(\partial F/\partial T\right)}_{V,N}=-S$.
Furthermore, Pareto law is no more valid for small incomes $I_i$ 
(for an empirical evidence see \cite{1} and references therein) and as a
consequence the integral in (\ref{40}) must be calculated with a lower bound,
i.e. $x_i\geq J > 0$. In this model we do not consider the variable $y_i$ 
in $I$.
Finally, under the condition $T<T_{max}$ ensuring the convergence
of the integral, we get:
\begin{eqnarray}
& &S=N\ln(JVT)+N-N\ln(T_{max}-T)+\frac{NT}{T_{max}-T},\label{41}\\
& &\frac{\overline{Y}}{N}=T+T_{max}\ln\left(\frac{J}{T_{max}}\right)+\frac{T^2}{T_{max}-T}.\label{42}
\end{eqnarray} 
For $T<<T_{max}$, expression (\ref{42}) becomes:
\begin{equation}
T=\frac{\overline{Y}}{N}-T_{max}\ln\left(\frac{J}{T_{max}}\right)+o(T).
\label{43}
\end{equation}
Note that in this case, conversely to the analogous free case with $M(x_i)$,
the temperature is not merely $\overline{Y}/N$.
Note that in general $J<T_{max}$. As a consequence the second terms in the right hand side
of (\ref{43}) gives a positive contribution to the temperature. 
Furthermore, as pointed in \cite{10}, the study of quantities 
such as $T\partial S/\partial T$ can give suitable information on
possible phase transitions of an economic system.\\
In our case, the calculation $T\partial S/\partial T$
indicates an economic transition phase
when $T=T_{max}$. In fact for $T=T_{max}$ the expressions for $\overline{Y},S,F,Z$ diverge.
This implies that the system can be in equilibrium only for $T<T_{max}$.  
This simple example shows how our formalism can give some predictions
concerning the possible macroeconomic changes of a given economic systems
representing a country.

\end{document}